\documentclass[smallextended,numbook]{svjour3_mod}

\usepackage[hidelinks]{hyperref}

\usepackage{amsmath}
\usepackage{amssymb}

\usepackage{mathptmx}
\usepackage{helvet}
\usepackage{courier}

\usepackage{graphicx}

\begin{document}

\title{The Laplace operator, measure concentration,\\
 Gauss functions, and quantum mechanics}

\author{Harry Yserentant}

\institute{
 Institut f\"ur Mathematik, Technische Universit\"at Berlin,
 10623 Berlin, Germany\\
\email{yserentant@math.tu-berlin.de}}

\date{August 8, 2022}

\titlerunning{The Laplace operator, measure 
 concentration, and quantum mechanics} 
 
\authorrunning{Harry Yserentant}

\maketitle

\noindent
{\bf Abstract}
\medskip

\noindent
We represent in this note the solutions of 
the electronic Schr\"odinger equation as traces 
of higher-dimensional functions. This allows 
to decouple the electron-electron interaction 
potential but comes at the price of a degenerate 
elliptic operator replacing the Laplace operator 
on the higher-dimensional space. The surprising 
observation is that this operator can without 
much loss again be substituted by the Laplace 
operator, the more successful the larger the 
system under consideration is. This is due 
to a concentration of measure effect that has  
much to do with the random projection theorem 
known from probability theory. The text is 
in parts based on the publications 
[Numer. Math. 146, 219--238 (2020)] and
[SIAM J. Matrix Anal. Appl., 43, 464--478 (2022)]
of the author and adapts the findings there 
to the needs of quantum mechanics. Our 
observations could for example find use in 
iterative methods that map sums of products 
of orbitals and geminals onto functions of 
the same type.

%
%
%

%
%
%
%
%
%
%
%
%


\renewcommand {\thefigure}{\arabic{figure}}

\newcommand   {\rmref}[1]   {{\rm (\ref{#1})}}

\newcommand   {\diff}[1]    {\mathrm{d}#1}

\newcommand   {\fourier}[1] {\widehat{#1}}

\newcommand   {\sector}[1]  {\lambda\big(\big\{x\,\big|\,#1\big\}\big)}

\def \dx      {\,\diff{x}}
\def \dy      {\,\diff{y}}

\def \deta    {\,\diff{\eta}}
\def \domega  {\,\diff{\omega}}

\def \dr      {\,\diff{r}}
\def \ds      {\,\diff{s}}
\def \dt      {\,\diff{t}}

\def \L       {\mathcal{L}}

\def \xy      {\bigg(\,\begin{matrix}x\\y\end{matrix}\,\bigg)}

\def \oe      {\bigg(\begin{matrix}\omega\\\eta\end{matrix}\bigg)}


\section{Introduction}
\label{sec1}

\setcounter{equation}{0}
\setcounter{theorem}{0}

The electronic Schr\"odinger equation establishes 
a connection between chemistry and physics. It 
describes systems of electrons that interact among 
each other and with a given, fixed set of nuclei. 
The electronic Schr\"odinger equation motivated 
our former work \cite{Yserentant_2020}, 
\cite{Yserentant_2022} centered around the 
approximate solution of high-dimensional partial 
differential equations. The present text compiles 
some of these results in view of applications in 
quantum theory and adapts and complements them 
correspondingly. We think here, for example, 
of procedures like the approximate inverse 
iteration 
\begin{equation}    \label{eq1.1}
u\;\,\leftarrow\;\, 
u\,-\,(-\Delta+\mu)^{-1}(-\Delta u+\mu u+Vu-\lambda(u)u)
\end{equation}
for the calculation of the ground state energy
or more general the first energy levels of a 
molecule, which requires the repeated 
(approximate) solution of the equation
\begin{equation}    \label{eq1.2}
-\Delta u+\mu u=f
\end{equation}
on the $\mathbb{R}^m$ for high dimensions $m$, where 
$\mu>0$ is a given constant. Provided the right-hand 
side $f$ of the equation (\ref{eq1.2}) possesses an 
integrable Fourier transform,  
\begin{equation}    \label{eq1.3}
u(x)=\bigg(\frac{1}{\sqrt{2\pi}}\bigg)^m\!\int
\frac{1}{\mu+\|\omega\|^2}\,\fourier{f}(\omega)\,
\mathrm{e}^{\,\mathrm{i}\,\omega\cdot x}\domega
\end{equation}
is a solution of this equation, and the only solution
that tends uniformly to zero as x goes to infinity.
If the right-hand side $f$ of the equation is a 
tensor product
\begin{equation}    \label{eq1.4}
f(x)=\prod_i\phi_i(x_i)
\end{equation}
of functions say from the three-dimensional 
space to the real numbers, or a sum of such 
tensor products, the same holds for the 
Fourier transform of $f$. If one replaces 
the corresponding term in the high-dimensional 
integral (\ref{eq1.3}) by an approximation 
\begin{equation}    \label{eq1.5}
\frac{1}{\mu+\|\omega\|^2}\approx
\sum_k a_k\,\mathrm{e}^{-\beta_k\left(\mu+\|\omega\|^2\right)}
= \sum_k a_k\,\mathrm{e}^{-\beta_k\mu}
\prod_i\mathrm{e}^{-\beta_k\|\omega_i\|^2}
\end{equation}
based on an approximation of $1/r$ by a sum 
of exponential functions, as described in 
Sect.~\ref{sec6}, for example, the integral 
collapses in this case to a sum of products 
of lower-dimensional integrals. That is, 
the solution can, independent of the space 
dimension, be approximated by a sum of 
such tensor products. The right-hand sides 
involved in iterations like (\ref{eq1.1})
are, however, not of such a simple structure. 
This is due to the fact that the potential 
$V$ is not only composed of terms depending 
on the position of a single electron in 
space, but also of terms that depend on the 
distance of two electrons. A corresponding 
ansatz for the solutions provided, this 
requires the solution of the equation 
(\ref{eq1.2}) with right-hand sides that 
are composed of terms of the form
\begin{equation}    \label{eq1.6}
f(x)=\bigg(\prod_i\phi_i(x_i)\bigg)
\bigg(\prod_{i<j}\phi_{ij}(x_i-x_j)\bigg).
\end{equation}
The question is whether this structure is 
reflected in the solution and the iterates stay 
in this class. To be precise, let us consider 
a system of $N$ electrons, let 
\begin{equation}    \label{eq1.7}    
m=3\times N, \quad n=3\times\frac{N(N+1)}{2}, 
\end{equation}
and let the vectors in $\mathbb{R}^m$ and 
$\mathbb{R}^n$, respectively, be partitioned into 
subvectors in the position space $\mathbb{R}^3$. 
Let the subvectors of the vectors in $\mathbb{R}^m$ 
and the first $N$ of the subvectors of the vectors 
in $\mathbb{R}^n$ be labeled by the indices 
$i=1,\ldots,N$ and the remaining subvectors 
of the vectors in $\mathbb{R}^n$ by the index 
pairs $(i,j)$, $i,j=1,\ldots, N$ and $i<j$. 
Let the right-hand side of the equation 
(\ref{eq1.2}) be of the form $f(x)=F(Tx)$, 
where  $T$ maps the vectors $x\in\mathbb{R}^m$ 
into the vectors $Tx\in\mathbb{R}^n$ with the 
subvectors
\begin{equation}    \label{eq1.8}
Tx|_i=x_i,\quad 
Tx|_{ij}=\frac{x_i-x_j}{\sqrt{2}},
\end{equation}
and where $F$ is a correspondingly structured 
function from the higher-dimensional space 
$\mathbb{R}^n$ to $\mathbb{R}$. If this function 
possesses an integrable Fourier transform,
the solution of the equation (\ref{eq1.2}) 
is the trace $u(x)=U(Tx)$ of the function
\begin{equation}    \label{eq1.9}
U(y)=\bigg(\frac{1}{\sqrt{2\pi}}\bigg)^n\!\int
\frac{1}{\mu+\|T^t\omega\|^2}\,\fourier{F}(\omega)\,
\mathrm{e}^{\,\mathrm{i}\,\omega\cdot y}\domega
\end{equation}
mapping the higher dimensional space to the 
real numbers. A in comparison to that in  
reference \cite{Yserentant_2020} considerably 
simplified, more direct proof of this proposition 
is given in Sect.~\ref{sec2}. The key result 
of the present paper is that the function
\begin{equation}    \label{eq1.10}
\widetilde{U}(y)=\bigg(\frac{1}{\sqrt{2\pi}}\bigg)^n\!\int
\frac{1}{\mu+\|\omega\|^2}\,\fourier{F}(\omega)\,
\mathrm{e}^{\,\mathrm{i}\,\omega\cdot y}\domega,
\end{equation}
that is, the solution of the equation
$-\Delta U+\mu U=F$ on the higher-dimensional
space, represents a with increasing particle 
number increasingly better and in the end 
almost perfect approximation of the function 
(\ref{eq1.9}), so that we arrived again at the 
case of right-hand sides of the form (\ref{eq1.4}). 
The reason is that the fraction of the vectors 
$\omega$ on the unit sphere for which the 
euclidean norm of $T^t\omega$ differs from 
one by more than a given small amount tends 
exponentially to zero as the number of 
electrons increases. This is a nontrivial 
concentration of measure phenomenon that has 
a lot to do with the random projection 
theorem (see Lemma~5.3.2 in \cite{Vershynin}, 
for example), which plays an important role in 
the data sciences and is in close connection 
with the Johnson-Lindenstrauss theorem
\cite{Johnson-Lindenstrauss}. Section~\ref{sec3} 
of this text, which is in large parts more or 
less directly taken from \cite{Yserentant_2022}, 
is devoted to the study of this effect.

This observation can unsurprisingly be used
in iterative methods for the solution of the 
Laplace-like equation (\ref{eq1.2}), or, as 
indicated above, for the calculation of the 
ground state of a molecular system by some 
kind of approximate inverse iteration. 
A~comprehensive theory of such methods 
(for the matrix case) is due to Knyazev and 
Neymeyr; see \cite{Knyazev-Neymeyr} and the 
references cited therein, and 
\cite{Rohwedder-Schneider-Zeiser} and  
\cite{Yserentant_2016} for the infinite 
dimensional case. But one can also speculate 
that this measure concentration effect allows 
to replace the original Schr\"odinger equation 
in case of sufficiently high particle numbers 
by a largely decoupled, higher-dimensional 
equation, with solutions whose traces become 
increasingly better approximations of the 
true wave functions.


\section{Solutions as traces of higher-dimensional functions}
\label{sec2}

\setcounter{equation}{0}
\setcounter{theorem}{0}
 
We are in the following mainly concerned with 
functions $U:\mathbb{R}^n\to\mathbb{R}$, 
$n$ a potentially high dimension, that possess 
a then also unique representation
\begin{equation}    \label{eq2.1}
U(y)=\bigg(\frac{1}{\sqrt{2\pi}}\bigg)^n
\!\int\fourier{U}(\omega)\,
\mathrm{e}^{\,\mathrm{i}\,\omega\cdot y}\domega
\end{equation}
in terms of an integrable function $\fourier{U}$, 
their Fourier transform. Such functions are by 
the Riemann-Lebesgue theorem uniformly continuous 
and vanish at infinity. The space $W_0(\mathbb{R}^n)$ 
of these functions becomes under the norm
\begin{equation}    \label{eq2.2}   
\|U\|=\bigg(\frac{1}{\sqrt{2\pi}}\bigg)^n\!\int
|\fourier{U}(\omega)|\domega
\end{equation}
a Banach space. Let $T$ be a still arbitrary 
$(n\times m)$-matrix of full rank $m<n$ and let
\begin{equation}    \label{eq2.3}
u:\mathbb{R}^m\to\mathbb{R}:x\to U(Tx)
\end{equation}
be the trace of a function in $U\in W_0(\mathbb{R}^n)$.
As the functions in $W_0(\mathbb{R}^n)$ are uniformly 
continuous, the same obviously holds for their traces. 
Because there is a constant~$c$ with $\|x\|\leq c\,\|Tx\|$
for all $x\in\mathbb{R}^m$, the trace functions 
(\ref{eq2.3}) vanish at infinity, too. The next lemma 
gives a criterion for the existence of partial 
derivatives of the trace functions, where we use 
the common multi-index notation. 

\begin{lemma}       \label{lm2.1}
Let $U:\mathbb{R}^n\to\mathbb{R}$ be a function 
in $W_0(\mathbb{R}^n)$ and let the functions
\begin{equation}    \label{eq2.4}
\omega\to (\mathrm{i}\,T^t\omega)^\beta\fourier{U}(\omega),
\quad \beta\leq\alpha,
\end{equation}
be integrable. The trace function \rmref{eq2.3} 
possesses then the partial derivative
\begin{equation}    \label{eq2.5}
(\mathrm{D}^\alpha u)(x)=
\bigg(\frac{1}{\sqrt{2\pi}}\bigg)^n
\!\int (\mathrm{i}\,T^t\omega)^\alpha\fourier{U}(\omega)\,
\mathrm{e}^{\,\mathrm{i}\,\omega\cdot Tx}\domega,
\end{equation}
which is like $u$ itself uniformly continuous 
and vanishes at infinity.
\end{lemma}

\begin{proof}
Let $e_k\in\mathbb{R}^m$ be the vector with the
components $e_k|_j=\delta_{kj}$. To begin with, 
we examine the limit behavior of the difference 
quotient
\begin{displaymath}
\frac{u(x+he_k)-u(x)}{h} =
\bigg(\frac{1}{\sqrt{2\pi}}\bigg)^n\!\int
\frac{\mathrm{e}^{\,\mathrm{i}\,h\omega\cdot Te_k}-1}{h}\;
\fourier{U}(\omega)\,
\mathrm{e}^{\,\mathrm{i}\,\omega\cdot Tx}\domega
\end{displaymath}
of the trace function as $h$ goes to zero. 
If the function 
$\omega\to\omega\cdot Te_k\,\fourier{U}(\omega)$
is integrable, this difference quotient tends, 
because of
\begin{displaymath}
\bigg|\,\frac{\mathrm{e}^{\,\mathrm{i}\,ht}-1}{h}\,\bigg|\leq\,|\,t\,|,
\quad
\lim_{h\to 0}\frac{\mathrm{e}^{\,\mathrm{i}\,ht}-1}{h}=\,\mathrm{i}\,t,
\end{displaymath}
by the dominated convergence theorem 
to the limit value
\begin{displaymath}
(\mathrm{D}_ku)(x) = 
\bigg(\frac{1}{\sqrt{2\pi}}\bigg)^n
\!\int\mathrm{i}\,\omega\cdot Te_k\,\fourier{U}(\omega)\,
\mathrm{e}^{\,\mathrm{i}\,\omega\cdot Tx}\domega.
\end{displaymath}
Because of $\omega\cdot Te_k=T^t\omega\cdot e_k$,
this proves (\ref{eq2.5}) for partial derivatives
of order one. For partial derivatives of higher
order, the proposition follows by induction.
\qed 
\end{proof}

Let $W_0^2(T)$ be the space of the functions 
$U\in W_0(\mathbb{R}^n)$ with finite (semi)-norm
\begin{equation}    \label{eq2.6}
|\,U|_T=\bigg(\frac{1}{\sqrt{2\pi}}\bigg)^n
\!\int\|T^t\omega\|^2\,|\fourier{U}(\omega)|\domega.
\end{equation}
The traces of these functions  are by Lemma~\ref{lm2.1} 
twice continuously differentiable. Let 
$\L:W_0^2(T)\to W_0(\mathbb{R}^n)$ be the degenerate 
elliptic differential operator given by
\begin{equation}    \label{eq2.7}
(\L U)(y)=\bigg(\frac{1}{\sqrt{2\pi}}\bigg)^n
\!\int\|T^t\omega\|^2\,\fourier{U}(\omega)\,
\mathrm{e}^{\,\mathrm{i}\,\omega\cdot y}\domega.
\end{equation}
For the functions $U\in W_0^2(T)$ and their 
traces (\ref{eq2.3}), by Lemma~\ref{lm2.1}
\begin{equation}    \label{eq2.8}
-\,(\Delta u)(x)=(\L U)(Tx)
\end{equation}
holds. The solutions of the equation (\ref{eq1.2})
thus are, with corresponding right-hand sides, the 
traces of the solutions $U\in W_0^2(T)$ of the 
differential equation
\begin{equation}    \label{eq2.9}
\L U+\mu U=F.
\end{equation}

\begin{theorem}     \label{thm2.1}
Let $F:\mathbb{R}^n\to\mathbb{R}$ be a function with 
integrable Fourier transform, let $f(x)=F(Tx)$, and 
let $\mu$ be a positive constant. The trace 
\rmref{eq2.3} of the function
\begin{equation}    \label{eq2.10}
U(y)=\bigg(\frac{1}{\sqrt{2\pi}}\bigg)^n\!\int
\frac{1}{\mu+\|T^t\omega\|^2}\,\fourier{F}(\omega)\,
\mathrm{e}^{\,\mathrm{i}\,\omega\cdot y}\domega
\end{equation}
is then twice continuously differentiable 
and the only solution of the equation
\begin{equation}    \label{eq2.11}
-\Delta u+\mu u=f
\end{equation}
whose values tend uniformly to zero
as $\|x\|$ goes to infinity.
\end{theorem}
That the trace $u$ is a classical solution 
of the equation (\ref{eq2.11}) follows from 
the remarks above, and that $u$ vanishes at 
infinity from the Riemann-Lebesgue theorem. 
By the maximum principle from the next 
lemma, the trace $u$ is the only such 
solution. 

\begin{lemma}       \label{lm2.2}
Let $u:\mathbb{R}^m\to\mathbb{R}$ be a twice
continuously differentiable function that 
vanishes at infinity. Let $\mu>0$ and let 
$-\Delta u+\mu u\geq 0$. Then $u\geq 0$ 
everywhere.
\end{lemma}

\begin{proof}
Let $u(x_0)<0$ at some point $x_0$ and let 
$|u(x)|<|u(x_0)|$ for $\|x\|\geq R$. Then there 
exists  a point $x_1$ of norm $\|x_1\|\leq R$ 
with $u(x_1)\leq u(x)$ for all~$x$ at first 
inside this ball. The function $u$ attains 
then at $x_1$ its global minimum. The Hessian 
of $u$ is at this point necessarily positive 
semidefinite, which implies that $\Delta u\geq 0$ 
there. Because $\mu>0$, at this point therefore 
$-\Delta u+\mu u<0$. This contradicts the 
assumption.
\qed
\end{proof}

As said, we are primarily interested in right-hand 
sides $F$ that themselves are products of 
lower-dimensional functions, or that are sums or 
rapidly converging series of such functions. 
It seems that the term
\begin{equation}    \label{eq2.12}
\frac{1}{\mu+\|T^t\omega\|^2}
\end{equation}
destroys this structure and that the transition
to the equation (\ref{eq2.9}) is therefore of 
little value. In high dimensions, however, 
often the contrary holds due to the concentration 
of measure effects studied in the next section. 
For the matrix (\ref{eq1.8}) assigned to the 
Schr\"odinger equation, the fraction of the 
vectors $\omega$ on the unit sphere of the 
$\mathbb{R}^n$ for which the euclidean norm 
of $T^t\omega$ differs from one by more than 
a given small amount tends exponentially to 
zero as the number of electrons goes to infinity. 
This means that we can replace the euclidean 
norm of $T^t\omega$ without much loss by that 
of $\omega$ itself and arrive again at the case 
described at the very beginning. If needed, the 
resulting approximations of the solutions of 
the equation (\ref{eq2.9}) can be iteratively 
improved.


\section{The underlying measure concentration effect}
\label{sec3}

\setcounter{equation}{0}
\setcounter{theorem}{0}

Let $n>m$ and let $A$ be a real $(m\times n)$-matrix
of rank $m$. The kernel of such a matrix has the 
dimension $n-m$ and hence can, in dependence of the 
dimensions, be a large subspace of the $\mathbb{R}^n$. 
Nevertheless, the set of all $x$ for which
\begin{equation}    \label{eq3.1}
\|Ax\|\geq\delta\,\|A\|\|x\|
\end{equation}
holds fills, in the high-dimensional case, often 
almost the complete $\mathbb{R}^n$ once $\delta$ 
falls below a certain bound; the norms are here 
and as in the previous section the euclidean norm 
on the $\mathbb{R}^m$ and the $\mathbb{R}^n$ and 
the assigned spectral norms of matrices. Let 
$\chi$ be the characteristic function of the 
set of all $x$ for which  
$\|Ax\|<\delta\,\|A\|\|x\|$ holds, and let 
$\nu_n$ be the volume of the unit ball in 
$\mathbb{R}^n$. The normed area measure
\begin{equation}    \label{eq3.2}
\frac{1}{n\nu_n}\int_{S^{n-1}}\!\chi(\eta)\deta
\end{equation}
of the subset of the unit sphere on which the 
condition (\ref{eq3.1}) is violated takes in 
such cases an extremely small value, which
conversely again means that (\ref{eq3.1}) holds 
on an overwhelmingly large part of the unit 
sphere and with that of the full space. We 
study this phenomenon in this section along 
the lines given in \cite{Yserentant_2022}
for orthogonal projections, matrices of the 
given kind with one as the only singular value, 
and for a class of matrices that in a sense 
do not substantially differ from such 
projections and under which the transpose of 
the matrix $T$ described in the introduction 
falls.

The surface integrals (\ref{eq3.2}) are not easily 
accessible and are difficult to calculate and 
estimate. We reformulate them therefore as volume 
integrals and draw some first conclusions from 
these representations. The starting point is the 
decomposition
\begin{equation}    \label{eq3.3}
\int_{\mathbb{R}^n}f(x)\dx\,= 
\int_{S^{n-1}}\bigg(\int_0^\infty\!f(r\eta)r^{n-1}\dr\bigg)\!\deta
\end{equation}
of the integrals of functions in $L_1$ into 
an inner radial and an outer angular part. 
Inserting the characteristic function of the
unit ball, one recognizes that the area of
the $n$-dimensional unit sphere is $n\nu_n$,
with $\nu_n$ the volume of the unit ball. If 
$f$ is rotationally symmetric, $f(r\eta)=f(re)$ 
holds for every $\eta\in S^{n-1}$ and every 
fixed, arbitrarily given unit vector $e$.
In this case, (\ref{eq3.3}) reduces 
therefore to
\begin{equation}    \label{eq3.4}
\int f(x)\dx\,=n\nu_n\int_0^\infty\!f(re)r^{n-1}\dr.
\end{equation}

\begin{lemma}       \label{lm3.1}
Let $A$ be an arbitrary matrix of dimension $m\times n$, 
$m<n$, let $\chi$ be the characteristic function of
the set of all $x\in\mathbb{R}^n$ for which 
$\|Ax\|<\delta\,\|A\|\|x\|$ holds, and let 
$W:\mathbb{R}^n\to\mathbb{R}$ be a rotationally 
symmetric weight function with integral
\begin{equation}    \label{eq3.5}
\int W(x)\dx\,=\,1.
\end{equation}
The weighted surface integral \rmref{eq3.2}  
then takes the value
\begin{equation}    \label{eq3.6}
\int \chi(x)W(x)\dx.
\end{equation}
\end{lemma}

\begin{proof}
Let $e$ be a given unit vector. For $\eta\in S^{n-1}$ and 
$r>0$ then $\chi(r\eta)=\chi(\eta)$ and $W(r\eta)=W(re)$ 
holds and the integral (\ref{eq3.6}) can by (\ref{eq3.3})
be written as
\begin{displaymath}
\int\chi(x)W(x)\dx\,= 
\int_{S^{n-1}}\chi(\eta)\bigg(\int_0^\infty\!W(re)r^{n-1}\dr\bigg)\!\deta.
\end{displaymath}
Because the inner integral takes by (\ref{eq3.4}) 
and (\ref{eq3.5}) the value
\begin{displaymath}
\int_0^\infty\!W(re)r^{n-1}\dr=\frac{1}{n\nu_n},
\end{displaymath}
this proves the proposition.
\qed
\end{proof}
An obvious choice for the weight function 
$W$ is the normed Gauss function
\begin{equation}    \label{eq3.7}
W(x)=\bigg(\frac{1}{\sqrt{\pi}}\bigg)^n\exp\big(-\|x\|^2\big).
\end{equation}
Another possible choice is the characteristic 
function of the unit ball $B$ of the 
$\mathbb{R}^n$ divided by its volume. For 
abbreviation, we introduce the probability 
measure
\begin{equation}    \label{eq3.8}
\lambda(M)=\frac{\mathrm{vol}(M\cap B)}{\mathrm{vol}(B)}
\end{equation}
on the measurable subsets $M$ of the 
$\mathbb{R}^n$.

\begin{lemma}       \label{lm3.2}
Let $A$ be a matrix of dimension $m\times n$, 
$m<n$. The weighted integral \rmref{eq3.2} 
over the surface of the unit ball is then 
equal to the volume ratio
\begin{equation}    \label{eq3.9}
\sector{\|Ax\|<\delta\,\|A\|\|x\|}.
\end{equation}
\end{lemma}

Because the euclidean length of a vector and 
the volume of a set are invariant to orthogonal 
transformations, the surface ratio (\ref{eq3.2}) 
and the volume ratio (\ref{eq3.9}) as well 
depend only on the singular values of the 
matrix under consideration.

\begin{lemma}       \label{lm3.3}
Let $A$ be a matrix of dimension $m\times n$, $m<n$, 
with singular value decomposition $A=U\Sigma V^t$.
The volume ratio \rmref{eq3.9} is then equal to 
the volume ratio
\begin{equation}    \label{eq3.10}
\sector{\|\Sigma x\|<\delta\,\|\Sigma\|\|x\|},
\end{equation}
that is, it depends exclusively on the singular 
values of the matrix $A$.
\end{lemma}

\begin{proof}
As the multiplication with the orthogonal matrices 
$U$ and $V^t$, respectively, does not change the 
euclidean norm of a vector, the set of all 
$x\in\mathbb{R}^n$ for which 
\begin{displaymath}
\|Ax\|<\delta\,\|A\|\|x\|,\quad \|x\|\leq R,
\end{displaymath}
holds coincides with the set of all $x$ for which 
we have
\begin{displaymath}
\|\Sigma V^tx\|<\delta\,\|\Sigma\|\|V^tx\|,
\quad \|V^tx\|\leq R.
\end{displaymath}
As the volume is invariant to orthogonal 
transformations, the proposition follows.
\qed
\end{proof}

Another simple and seemingly obvious observation
is the following lemma.

\begin{lemma}       \label{lm3.4}
Let $A$ be a matrix of dimension $m\times n$, 
$m<n$, of full rank $m$. Then
\begin{equation}    \label{eq3.11}
\lim_{\delta\to 1-}\sector{\|Ax\|<\delta\,\|A\|\|x\|}\,=\,1.
\end{equation}
\end{lemma}

\begin{proof}
By Lemma~\ref{lm3.3}, we can restrict ourselves 
to diagonal matrices $A=\Sigma$. Because the 
limit takes by the dominated convergence theorem
the value
\begin{displaymath}
\sector{\|\Sigma x\|<\|\Sigma\|\|x\|}
\end{displaymath}
and since the set of all $x$ for which 
$\|\Sigma x\|=\|\Sigma\|\|x\|$ holds is as 
a lower-dimensional subspace of the 
$\mathbb{R}^n$ a set of measure zero, the 
proposition follows.
\qed
\end{proof}

For orthogonal projections, matrices $A$ with 
one as the only singular value, the volume 
ratios (\ref{eq3.9}) possess a closed integral 
representation.

\begin{theorem}     \label{thm3.1}
Let the $(m\times n)$-matrix $P$ be an 
orthogonal projection. Then
\begin{equation}    \label{eq3.12}
\sector{\|Px\|<\delta\,\|x\|}=F(\delta), 
\quad 0\leq\delta<1,
\end{equation}
holds, where the function $F(\delta)=F(m,n;\delta)$ 
is defined by the integral expression
\begin{equation}    \label{eq3.13}
F(\delta)=
\frac{2\,\Gamma(n/2)}{\Gamma(m/2)\Gamma((n-m)/2)}\,
\int_0^\delta(1-t^2)^\alpha t^{m-1}\dt
\end{equation}
and the exponent $\alpha\geq -1/2$ is given by
\begin{equation}    \label{eq3.14}
\alpha=\frac{n-m-2}{2}.
\end{equation}
It takes nonnegative values for dimensions
$n\geq m+2$.
\end{theorem}

\begin{proof}
By Lemma~\ref{lm3.3}, we can restrict ourselves
to the matrix $P$ that extracts from a vector
in $\mathbb{R}^n$ its first $m$ components.
Consistent within the proof, we split the vectors 
in $\mathbb{R}^n$ into parts $x\in\mathbb{R}^m$ 
and $y\in\mathbb{R}^{n-m}$. The set whose volume 
has to be calculated consists then of the 
points in the unit ball for which
\begin{displaymath}
\|x\|<\delta\,\bigg\|\xy\bigg\|
\end{displaymath}
or, resolved for the norm of the component
$x\in\mathbb{R}^m$,
\begin{displaymath}
\|x\|<\varepsilon\,\|y\|, \quad
\varepsilon=\frac{\delta}{\sqrt{1-\delta^2}},
\end{displaymath}
holds. The volume can then be expressed 
as double integral 
\begin{displaymath}
\int\bigg(\int H\big(\varepsilon\|y\|-\|x\|\big)
\chi\big(\|x\|^2+\|y\|^2\big)\dx\bigg)\dy,
\end{displaymath}
where $H(t)=0$ for $t\leq0$, $H(t)=1$ for $t>0$,
$\chi(t)=1$ for $t\leq 1$, and $\chi(t)=0$ for 
arguments $t>1$. In terms of polar coordinates, 
that is, by (\ref{eq3.4}), it reads as
\begin{displaymath}
(n-m)\nu_{n-m}\int_0^\infty\bigg(
m\nu_m\int_0^{\varepsilon s}\chi\big(r^2+s^2\big)r^{m-1}\dr
\bigg)s^{n-m-1}\ds,
\end{displaymath}
with $\nu_d$ the volume of the $d$-dimensional unit 
ball. Substituting $t=r/s$ in the inner integral,
the upper bound becomes independent of $s$ and 
the integral can be written~as
\begin{displaymath}
(n-m)\nu_{n-m}\int_0^\infty\bigg(m\nu_m\,s^m\!
\int_0^{\varepsilon}\chi\big(s^2(1+t^2)\big)t^{m-1}\dt
\bigg)s^{n-m-1}\ds.
\end{displaymath}
Interchanging the order of integration, 
it attains the value
\begin{displaymath}
\frac{(n-m)\nu_{n-m}\,m\nu_m}{n}
\int_0^{\varepsilon}\frac{t^{m-1}}{(1+t^2)^{n/2}}\dt.
\end{displaymath}
For abbreviation, we introduce the function 
\begin{displaymath}
f(\delta)=g\bigg(\frac{\delta}{\sqrt{1-\delta^2}}\bigg),
\quad
g(\varepsilon)=
\int_0^{\varepsilon}\frac{t^{m-1}}{(1+t^2)^{n/2}}\dt,
\end{displaymath}
on the interval $0\leq\delta<1$. With $\alpha$
given by (\ref{eq3.14}), its derivative is
\begin{displaymath}
f'(\delta)=(1-\delta^2)^\alpha\delta^{m-1}.
\end{displaymath}
Because $f(0)=0$, it possesses therefore the 
representation
\begin{displaymath}
f(\delta)=\int_0^\delta(1-t^2)^\alpha t^{m-1}\dt.
\end{displaymath}
Dividing the expression above by $\nu_n$ and 
remembering that
\begin{displaymath}
\nu_d=\frac{2}{d}\,\frac{\pi^{d/2}}{\Gamma(d/2)},
\end{displaymath}
this completes the proof of the theorem.
\qed
\end{proof}
If the difference of the dimensions is even, the 
function (\ref{eq3.13}) is an either even or odd 
polynomial of degree $n-2$ in $\delta$. A closed 
representation is given in \cite{Yserentant_2022}. 
For practical purposes, it is more advantageous 
to calculate $F(\delta)$ numerically by means
of a quadrature rule. By (\ref{eq3.11}), $F$ 
takes the value $F(1)=1$. Thus
\mbox{$F(\delta)=F(\delta)/F(1)$}, so that 
there is no need to evaluate the Gamma function.

The function (\ref{eq3.13}) always represents 
a lower bound for the volume ratios (\ref{eq3.9}),
independent of particular properties of the 
matrix under consideration.

\begin{theorem}     \label{thm3.2}
Let $A$ be a nonvanishing matrix of dimension $m\times n$, 
$m<n$. Then
\begin{equation}    \label{eq3.15}
\sector{\|Ax\|<\delta\,\|A\|\|x\|}\,\geq\,F(m,n;\delta).
\end{equation}
\end{theorem}

\begin{proof}
We can restrict ourselves to diagonal matrices 
$A=\Sigma$. Let $P$ be the matrix that extracts 
from a vector in $\mathbb{R}^n$ the first $m$ 
components.
As $\|\Sigma x\|\leq\|\Sigma\|\|Px\|$ and 
$\|\Sigma\|>0$, the given volume ratio is
then bounded from below by the volume ratio
\begin{displaymath}  
\sector{\|Px\|<\delta\,\|x\|}.
\end{displaymath}
The proposition thus follows from 
Theorem~\ref{thm3.1}.
\qed
\end{proof}

Upper bounds for the volume ratio (\ref{eq3.9})
depend in general on the singular values of
the matrix, in the extreme case on its condition 
number, the ratio of its maximum and its minimum 
singular value \cite{Yserentant_2022}. This is 
fortunately not the case for the matrices 
$A=T^t$, with $T$ given by (\ref{eq1.8}), in 
which we are interested here and to which the 
next theorem applies. The euclidean norm of 
the vector $Tx\in\mathbb{R}^n$ is given by
\begin{equation}    \label{eq3.16}
\|Tx\|^2=\,\sum_{i=1}^N\|x_i\|^2+\,
\frac14\,\sum_{i=1}^N\sum_{j=1}^N\|x_i-x_j\|^2
\end{equation}
or, after rearrangement, with the rank 
three map $T_0x=x_1+x_2+\cdots+x_N$ by
\begin{equation}    \label{eq3.17}
\|Tx\|^2=\,\frac{N+2}{2}\,\|x\|^2-\,\frac12\,\|T_0x\|^2.
\end{equation}
The $(m\times m)$-matrix $T^tT$ thus has the 
eigenvalue $(N+2)/2$ of multiplicity $m-3$ and 
the eigenvalue $1$ of multiplicity $3$. The 
singular values of the matrix $T^t$ are 
therefore
\begin{equation}    \label{eq3.18}
\sigma_i=1\;\text{for $\,i\leq 3$},\quad
\sigma_i=\sqrt{\frac{N+2}{2}}\;\,\text{for $\,i\geq 4$}.
\end{equation}
The spectral norm of the matrix $T^t$ 
is $\sqrt{(N+2)/2}$.

\begin{theorem}     \label{thm3.3}
Let $n>m$ and let $A$ be a nonvanishing 
$(m\times n)$-matrix with singular values 
$\sigma_k=\sigma_m$ for $k>m_0$. The 
volume ratio \rmref{eq3.9} satisfies 
then the estimate
\begin{equation}    \label{eq3.19}
\sector{\|Ax\|<\delta\,\|A\|\|x\|}\leq F(m-m_0,n;\delta).
\end{equation}
\end{theorem}

\begin{proof}
We can restrict ourselves to diagonal matrices 
$A=\Sigma$. Let $P'$ be the matrix that 
extracts from a vector $x\in\mathbb{R}^n$ its 
components $x_k$, $m_0<k\leq m$. As
$\|\Sigma\|\|P'x\|\leq\|\Sigma x\|$ and
$\|\Sigma\|>0$, the volume ratio \rmref{eq3.19} 
is then less than or equal to the volume ratio
\begin{displaymath}    
\sector{\|P'x\|<\delta\,\|x\|}. 
\end{displaymath}
The proposition follows again from its 
representation in Theorem~\ref{thm3.1}.
\qed
\end{proof}

Next, we study the limit behavior of the 
function (\ref{eq3.13}) when the dimensions 
tend to infinity. The subsequent estimates 
are expressed in terms of the function
\begin{equation}    \label{eq3.20}
\phi(\vartheta) =
\vartheta\,\exp\bigg(\frac{1-\vartheta^2}{2}\bigg).
\end{equation}
It increases on the interval $0\leq\vartheta\leq 1$ 
strictly, attains at the point $\vartheta=1$ its 
maximum value one, and decreases from there again 
strictly.

\begin{theorem}     \label{thm3.4}
Let~$\xi$ be the square root of the dimension 
ratio $m/n$. For $\delta<\xi$, then 
\begin{equation}    \label{eq3.21}
0\,\leq\,F(m,n;\delta)
\leq\,\phi\bigg(\frac{\delta}{\xi}\bigg)^m.
\end{equation}
\end{theorem}

\begin{proof}
Let $P$ be the matrix that extracts from 
a vector in $\mathbb{R}^n$ its first $m$ 
components. The characteristic function 
$\chi$ of the set of all $x$ for which 
\mbox{$\|Px\|<\delta\,\|x\|$} holds 
satisfies, for any $t>0$, the crucial, 
even if obviously not very sharp estimate
\begin{displaymath}      
\chi(x)<
\exp\big(\,t\,\big(\delta^2\|x\|^2-\|Px\|^2\big)\big)
\end{displaymath}  
by a product of univariate functions.
By Lemma~\ref{lm3.1} and the subsequent 
remark, the volume ratio (\ref{eq3.12}) 
can therefore be estimated by the 
integral
\begin{displaymath}      
\bigg(\frac{1}{\sqrt{\pi}}\bigg)^n\!\int
\exp\big(\,t\,\big(\delta^2\|x\|^2-\|Px\|^2\big)\big)
\exp\big(-\|x\|^2\big)\dx
\end{displaymath}    
that remains finite for all $t$ in the 
interval $0\leq t<1/\delta^2$. It splits 
into a product of one-dimensional integrals 
and takes, for given $t$, the value 
\begin{displaymath}
\bigg(\frac{1}{1-\delta^2 t+t}\bigg)^{m/2}
\bigg(\frac{1}{1-\delta^2 t}\bigg)^{(n-m)/2}.
\end{displaymath}
This expression attains its minimum on 
the interval $0<t<1/\delta^2$ at
\begin{displaymath}
t=\frac{\xi^2-\delta^2}{(1-\delta^2)\delta^2}
\end{displaymath}
and takes at this point $t$ the value
\begin{displaymath}
\bigg(\,\frac{\delta}{\xi}\,
\bigg(\frac{1-\delta^2}{1-\xi^2}\bigg)^\gamma\;
\bigg)^m,  \quad \gamma=\frac{1-\xi^2}{2\xi^2}.
\end{displaymath}
If one sets $\delta/\xi=\vartheta$ for 
abbreviation, the logarithm
\begin{displaymath}
\ln\bigg(\bigg(\frac{1-\delta^2}{1-\xi^2}\bigg)^\gamma\;\bigg)
=\,\frac{1-\xi^2}{2\xi^2}\,
\ln\bigg(\frac{1-\vartheta^2\xi^2}{1-\xi^2}\bigg)
\end{displaymath}
possesses, because of $\vartheta^2\xi^2<1$ 
and $\xi^2<1$, the power series expansion
\begin{displaymath}
\frac {1-\vartheta^2}{2}-\frac12\,\sum_{k=1}^\infty
\bigg(\frac{1-\vartheta^{2k}}{k}-\,\frac{1-\vartheta^{2k+2}}{k+1}\bigg)\xi^{2k}.
\end{displaymath}
Because the series coefficients are for all 
$\vartheta\geq 0$ greater than or equal to 
zero and, by the way, polynomial multiples of 
$(1-\vartheta^2)^2$, the proposition follows. 
\qed
\end{proof}

Theorem~\ref{thm3.4} possesses a counterpart 
that deals with values $\delta$ greater 
than the square root of the ratio of the 
dimensions $m$ and $n$.

\begin{theorem}     \label{thm3.5}
Let~$\xi$ be the square root of the dimension 
ratio $m/n$. For $\delta>\xi$, then 
\begin{equation}    \label{eq3.22}
0\,\leq\,1-F(m,n;\delta)
\leq\,\phi\bigg(\frac{\delta}{\xi}\bigg)^m.
\end{equation}
\end{theorem}

\begin{proof}
The proof almost coincides with that of 
the previous theorem. Let $P$ be again 
the matrix that extracts from a vector 
in $\mathbb{R}^n$ its first $m$ 
components. Instead of the volume 
ratio (\ref{eq3.12}), now we have to 
estimate the volume ratio
\begin{displaymath} 
1-\sector{\|Px\|<\delta\,\|x\|} =
\sector{\|Px\|\geq\delta\,\|x\|}.
\end{displaymath}
For sufficiently small positive values $t$, 
it can be estimated  by the integral
\begin{displaymath}
\bigg(\frac{1}{\sqrt{\pi}}\bigg)^n\!\int
\exp\big(\,t\,\big(\|Px\|^2-\delta^2\|x\|^2\big)\big)
\exp\big(-\|x\|^2\big)\dx.
\end{displaymath}
This integral splits into a product of 
one-dimensional integrals and takes 
the value
\begin{displaymath}
\bigg(\frac{1}{1+\delta^2 t-t}\bigg)^{m/2}
\bigg(\frac{1}{1+\delta^2 t}\bigg)^{(n-m)/2},
\end{displaymath}
which attains, for $\delta<1$, on the interval 
$0<t<1/(1-\delta^2)$ its minimum at
\begin{displaymath}
t=\frac{\delta^2-\xi^2}{(1-\delta^2)\delta^2}. 
\end{displaymath}
It takes at this point $t$ again the value
\begin{displaymath}
\bigg(\,\frac{\delta}{\xi}\,
\bigg(\frac{1-\delta^2}{1-\xi^2}\bigg)^\gamma\;
\bigg)^m,  \quad \gamma=\frac{1-\xi^2}{2\xi^2}.
\end{displaymath}
This leads as in the proof of Theorem~\ref{thm3.4} 
to the estimate (\ref{eq3.22}). 
\qed
\end{proof}

\begin{figure}[t]   \label{fig1}
\includegraphics[width=0.93\textwidth]{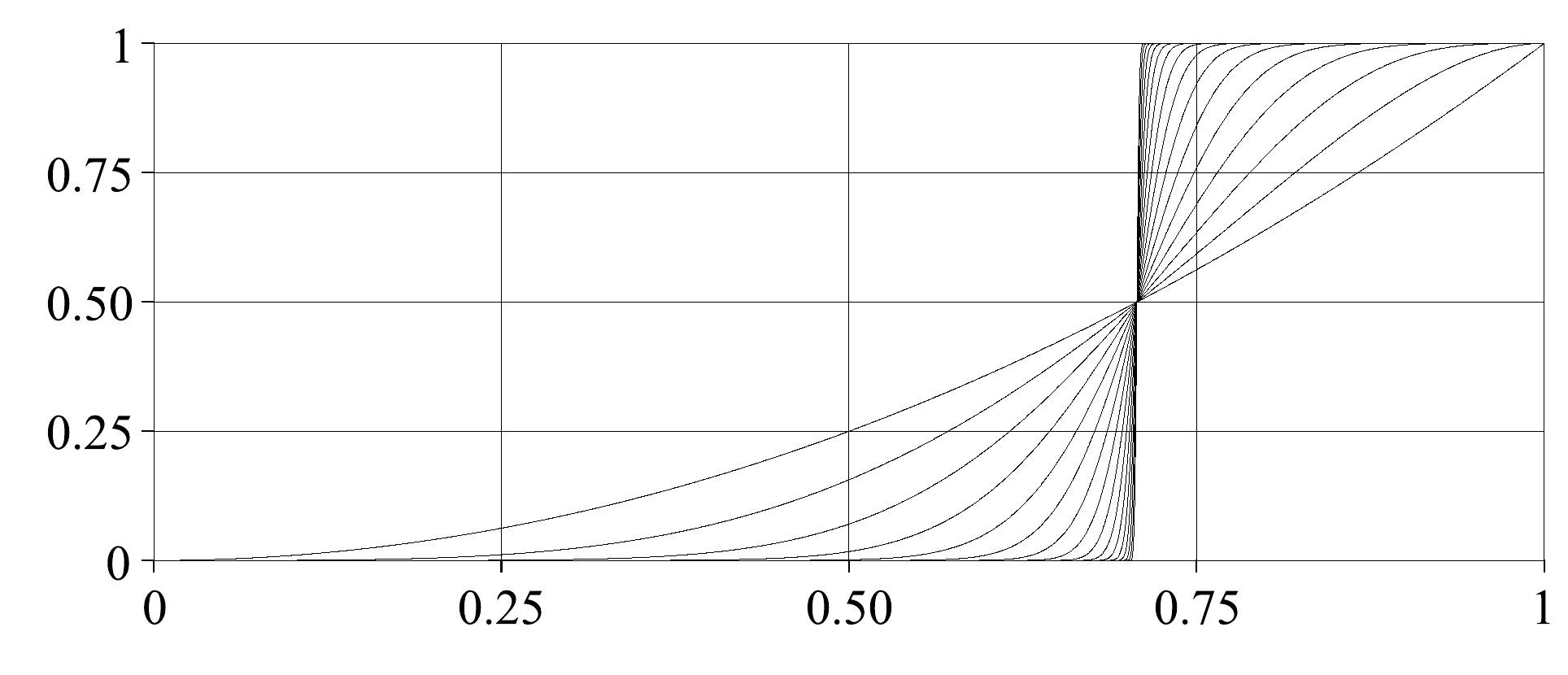}
\caption{The volume ratio \rmref{eq3.12} as function 
 of $0\leq\delta<1$ for $m=2^k$, $k=1,\ldots,16$, 
 and $n=2m$}
\end{figure}

If the dimension ratio $\delta_0^2=m/n$ is kept 
fixed or only tends to $\delta_0^2$, the volume 
ratios (\ref{eq3.12}) and the functions 
(\ref{eq3.13}), respectively, thus tend 
to a step function with jump discontinuity at 
$\delta_0$. Figure~1 reflects this behavior.
We summarize our findings therefore once more 
as follows and relate them to the prospective 
jump positions.

\begin{theorem}     \label{thm3.6}
Let $A$ be a nonvanishing matrix of dimension $m\times n$, 
$m<n$, and let $\xi$ be the square root of the dimension 
ratio $m/n$. For $\vartheta>1$ then one has
\begin{equation}    \label{eq3.23}
\sector{\|Ax\|\geq\vartheta\xi\|A\|\|x\|}\leq\phi(\vartheta)^m.
\end{equation}
\end{theorem}

\begin{proof}
As the set of the $x$ for which $\|Ax\|\geq\|A\|\|x\|$ 
holds is as a lower-dimensional subspace of the 
$\mathbb{R}^n$ a set of measure zero, we can restrict 
ourselves to the case $\vartheta\xi<1$. Setting 
$\delta=\vartheta\xi$, the proposition then follows 
from Theorem~\ref{thm3.2} and Theorem~\ref{thm3.5}. 
\qed
\end{proof}

The theorem states in particular that the norm of 
$Ax$ exceeds the value $\xi\|A\|\|x\|$ by more than 
a moderate factor $\vartheta>1$ only on a very small, 
de facto negligibly sector, an observation that is 
of great importance for the analysis of iterative 
methods for the solution of equations like (\ref{eq2.9}). 
Under the much more restrictive assumptions from 
Theorem~\ref{thm3.3}, Theorem~\ref{thm3.6} possesses 
a counterpart for values $\vartheta<1$.

\begin{theorem}     \label{thm3.7}
Let $n>m$ and let $A$ be a nonvanishing $(m\times n)$-matrix  
with singular values $\sigma_k=\sigma_m$ for $k>m_0$. If 
$m'=m-m_0$ and $\xi'$ is the square root of $m'/n$, then
\begin{equation}    \label{eq3.24}
\sector{\|Ax\|<\vartheta\xi'\|A\|\|x\|}\leq\phi(\vartheta)^{m'}
\end{equation}
holds for all $\vartheta$ in the interval $0<\vartheta<1$.
\end{theorem}

\begin{proof}
This results from the estimates in Theorem~\ref{thm3.3} 
and Theorem~\ref{thm3.4}.
\qed
\end{proof}


\section{Back to the equation}
\label{sec4}

\setcounter{equation}{0}
\setcounter{theorem}{0}

\setcounter{figure}{1}

The last two theorems apply, because of (\ref{eq3.18}), 
to the matrices $A=T^t$ assigned to the Schr\"odinger 
equation. They form the basis of our argumentation. 
For a randomly chosen vector $\omega$ in the frequency 
or momentum space, the probability that
\begin{equation}    \label{eq4.1}
(1-\varepsilon)\xi'\,\|T^t\|\|\omega\|\leq\|T^t\omega\|
<(1+\varepsilon)\xi\,\|T^t\|\|\omega\|
\end{equation}
holds for a given $\varepsilon$ between $0$ 
and $1$ is by these theorems at least 
\begin{equation}    \label{eq4.2}
1-\big(\phi(1-\varepsilon)^{m'}+\phi(1+\varepsilon)^m\big),
\end{equation}
where $\phi$ is the function (\ref{eq3.20}), 
the dimensions 
\begin{equation}    \label{eq4.3}    
m=3\times N, \quad n=3\times\frac{N(N+1)}{2},
\end{equation} 
and $m'=3\times(N-1)$ depend on the number 
$N$ of particles, $\xi$ and $\xi'$ are the 
square roots of the dimension ratios $m/n$ 
and $m'/n$, and the constants 
\begin{equation}    \label{eq4.4}
\xi'\,\|T^t\|=\sqrt{1-\frac{2}{N(N+1)}}, 
\quad
\xi\,\|T^t\|=\sqrt{\vphantom{1-\frac{2}{N(N+1)}}1+\frac{1}{N+1}}
\end{equation}
enclose the value one and tend to one as $N$ goes 
to infinity. The norm of $T^t\omega$ thus rapidly 
approaches that of $\omega$ when the number of 
particles increases. Because of
\begin{equation}    \label{eq4.5}
\phi(1\pm\varepsilon)<\exp(-c\,\varepsilon^2),
\quad c=-\ln(\phi(2)),
\end{equation}
for values $0<\varepsilon<1$, the probability that 
(\ref{eq4.1}) holds is in any case greater than 
\begin{equation}    \label{eq4.6}
1-2\exp(-c\,\varepsilon^2m').
\end{equation}
A less easily interpretable, but significantly 
better lower bound than (\ref{eq4.2}) can be 
derived from Theorem~\ref{thm3.2} and 
Theorem~\ref{thm3.3}. In terms of the function 
(\ref{eq3.13}), it reads
\begin{equation}    \label{eq4.7}
F(m,n;(1+\varepsilon)\xi)-F(m',n;(1-\varepsilon)\xi')
\end{equation}
and deviates for increasing particle number 
less and less from the probability
\begin{equation}    \label{eq4.8}
F(m,n;(1+\varepsilon)\xi)-F(m,n;(1-\varepsilon)\xi)
\end{equation}
that an orthogonal projection from the $\mathbb{R}^n$
to the $\mathbb{R}^m$ maps a randomly chosen unit 
vector to a vector of length between $(1-\varepsilon)\xi$ 
and $(1+\varepsilon)\xi$. Because the products of the 
matrix $T^t$ with vectors $e=e_i$ and $e=e_{ij}$ in 
$\mathbb{R}^n$ pointing into the direction of the 
coordinate spaces have the norm $\|T^te\|=\|e\|$, 
they satisfy the condition
\begin{equation}    \label{eq4.9}
\xi'\,\|T^t\|\|e\|<\|T^te\|<\xi\,\|T^t\|\|e\|.
\end{equation}
This establishes a link to hyperbolic cross 
spaces and not least to the mixed regularity 
of electronic wave functions
\cite{Yserentant_2010}, \cite{Yserentant_2011}.

\begin{figure}[t]   \label{fig2}
\includegraphics[width=0.93\textwidth]{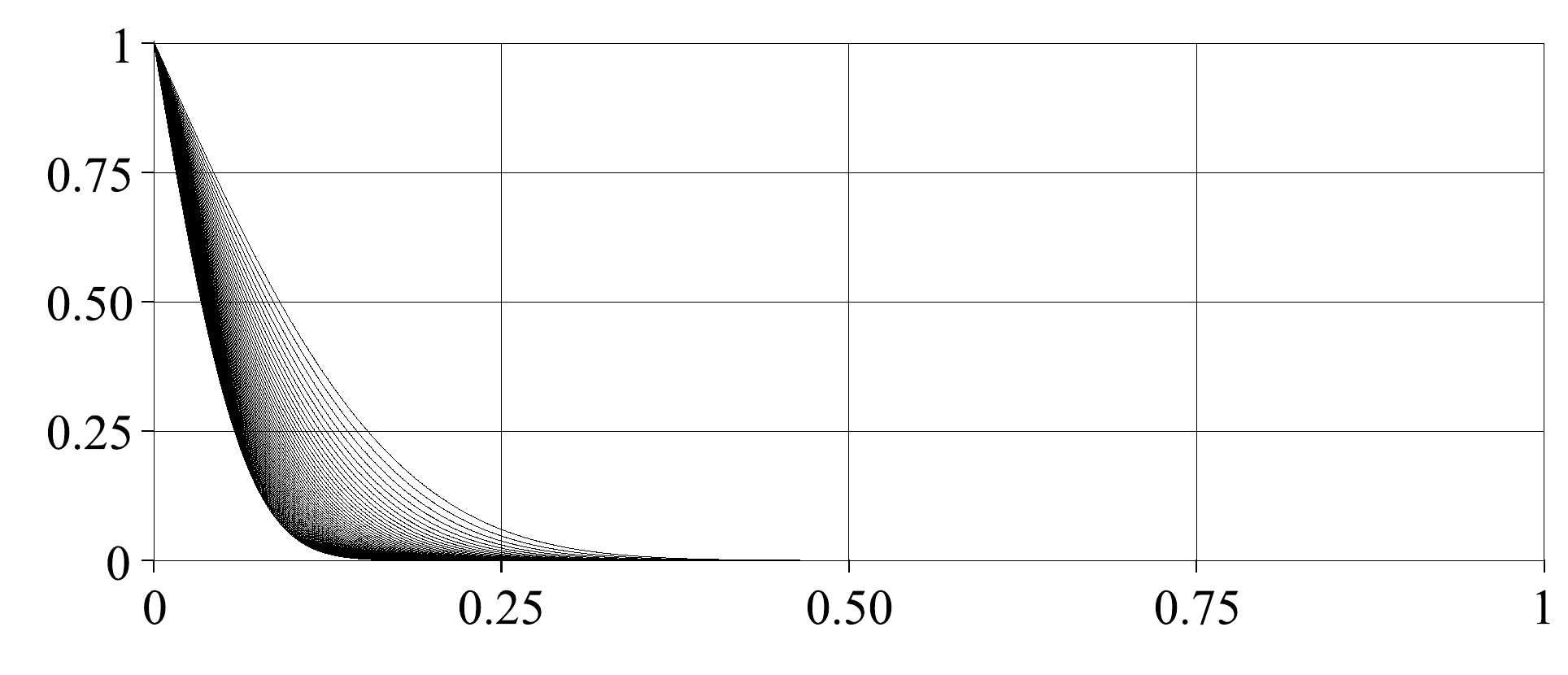}
\caption{The distance of the lower bound 
 \rmref{eq4.7} to one as function of 
 $\varepsilon$ for $N=8,\ldots,64$ electrons.}
\end{figure}

What does all this mean? For high electron numbers, 
at the latest when statistical physics comes into 
play, the norm of $T^t\omega$ is almost equal to 
that of $\omega$ for all $\omega$ outside of a tiny, 
probably largely negligible sector. Maybe that 
this allows to replace the operator (\ref{eq2.7}) 
in such cases by the negative Laplace operator and 
the original Hamilton operator to some extent by 
a correspondingly decoupled operator. But also 
for moderate particle numbers, the function 
(\ref{eq1.10}),
\begin{equation}    \label{eq4.10}  
\widetilde{U}(y)=  
\bigg(\frac{1}{\sqrt{2\pi}}\bigg)^n\!\int
\frac{1}{\mu+\|\omega\|^2}\,\fourier{F}(\omega)\,
\mathrm{e}^{\,\mathrm{i}\,\omega\cdot y}\domega,
\end{equation}
that is, the solution of the equation 
$-\Delta U+\mu U=F$ in the higher-dimensional 
space, remains a good approximation to the 
solution (\ref{eq2.10}) of the equation 
(\ref{eq2.9}), surely good enough to serve 
as a building block for the construction of 
rapidly convergent iterative methods. Figure~2 
shows the distance of the bound (\ref{eq4.7}) 
to one as function of $\varepsilon$ for some 
small to medium size systems.


\section{Square integrable right-hand sides}
\label{sec5}

\setcounter{equation}{0}
\setcounter{theorem}{0}

\setcounter{figure}{2}

The fact that the function (\ref{eq4.10}) is a good 
approximation of the solution (\ref{eq2.10}) of the 
equation (\ref{eq2.9}) depends in no way on the 
integrability of the Fourier transform of the 
right-hand side. The problem is that, for functions 
in the majority of $L_2$-based spaces, the transition 
to the trace functions is much more delicate than 
for the functions with integrable Fourier transforms. 
To a certain extent, smoothing of the right-hand 
side helps and mitigates this problem. Let 
$K:\mathbb{R}^n\to\mathbb{R}$ be a both integrable 
and square integrable normed kernel, like one of 
those with the Fourier transforms 
\begin{equation}    \label{eq5.1}
\fourier{K}(\omega)=
\bigg(\frac{1}{\sqrt{2\pi}}\bigg)^n
\prod_{i=1}^n\phi\bigg(\frac{\omega_i^2}{2}\bigg),
\quad \phi(t)=
\mathrm{e}^{-t}\sum_{k=0}^\nu\frac{t^k}{k!},
\end{equation}
depicted in Fig.~3 and with vanishing 
moments up to order $2\nu+1$, and let 
\begin{equation}    \label{eq5.2}
K_\varepsilon(x)=
\frac{1}{\varepsilon^n}\,K\Big(\frac{x}{\varepsilon}\Big).
\end{equation}
The convolution $K_\varepsilon*F$ both of square 
integrable functions $F$ and of functions $F$ 
with integrable Fourier transform possesses then 
the Fourier transform
\begin{equation}    \label{eq5.3}   
(\fourier{K_\varepsilon*F})(\omega)=
(\sqrt{2\pi})^n\fourier{K}(\varepsilon\omega)\fourier{F}(\omega), 
\end{equation}
which is in both cases integrable. For functions 
$F$ with integrable Fourier transform, this follows 
from the boundedness of $\fourier{K}$, and for 
square integrable functions $F$, it follows from the 
Cauchy-Schwarz inequality and Plancherel's theorem.

\begin{figure}[t]   \label{fig3}
\includegraphics[width=0.93\textwidth]{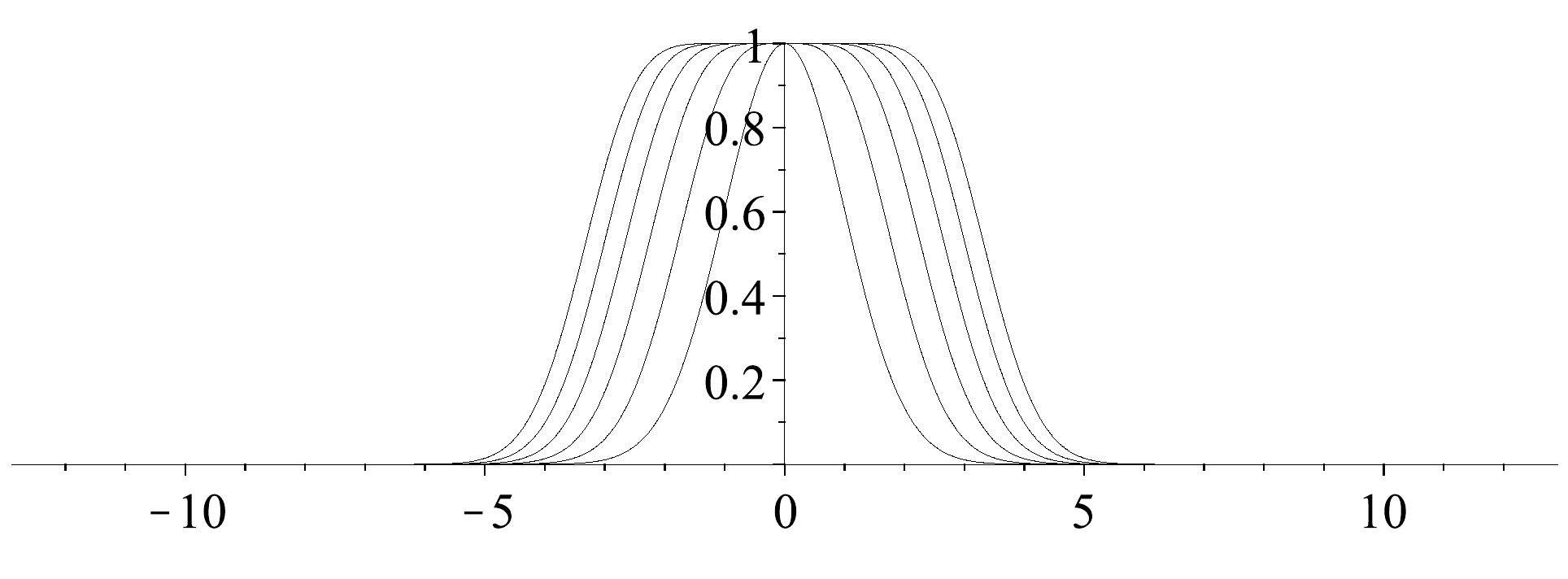}
\caption{The renormalized Fourier transforms 
 \rmref{eq5.1}  in one space dimension for 
 $\nu=0,1,\ldots,5$}
\end{figure}

The transition from the right-hand side $F$ to 
its smoothed variant $K_\varepsilon*F$ means 
that the solution $U$ of the equation 
(\ref{eq2.9}) is replaced by its smoothed 
counterpart $K_\varepsilon*U$. If the Fourier 
transform of $F$ is integrable, not much gets 
lost. The reason is that the maximum norm 
of functions in $W_0(\mathbb{R}^n)$ can be 
estimated by the scaled $L_1$-norm 
(\ref{eq2.2}) of their Fourier transforms. 
As $\varepsilon$ goes to zero, the functions 
$K_\varepsilon*U$ converge therefore 
uniformly to $U$ and their traces 
$u_\varepsilon$ uniformly to the trace $u$ 
of $U$. Because of their representation from 
Lemma~\ref{lm2.1}, the first- and second 
order derivatives of the $u_\varepsilon$ 
converge in the same way uniformly to the 
corresponding derivatives of~$u$. For square 
integrable right-hand sides $F$, the situation 
is considerably more complicated. Only if the 
traces of the functions $K_\varepsilon*F$ 
tend in the $L_2$-sense to a limit function $f$, 
the $u_\varepsilon$ tend in the Sobolev space 
$H^2$ to a solution $u$ of the equation 
(\ref{eq2.11}).


\section{The approximation by exponential functions}
\label{sec6}

\setcounter{equation}{0}
\setcounter{theorem}{0}

\setcounter{figure}{3}

We still need to approximate $1/r$ on given intervals 
$\mu\leq r\leq R\mu$ with moderate to high relative 
accuracy by sums of exponential functions. It suffices 
to restrict oneself to the case $\mu=1$, that is, to 
intervals $1\leq r\leq R$. If $v(r)$ approximates 
$1/r$ on this interval with a given relative 
accuracy, the rescaled function
\begin{equation}    \label{eq6.1}
r\;\to\;\frac{1}{\mu}\,v\bigg(\frac{r}{\mu}\bigg)
\end{equation}
approximates $1/r$ on the interval $\mu\leq r\leq R\mu$ 
with the same relative accuracy. Our favorite 
approximations of $1/r$ are the at first sight harmless
looking sums
\begin{equation}    \label{eq6.2}
v(r)=h\sum_{k=k_1}^{k_2}
\mathrm{e}^{-kh}\exp(-\mathrm{e}^{-kh}r),
\end{equation}
a construction that is due to Beylkin and Monz\'{o}n
\cite{Beylkin-Monzon}. The parameter $h$ determines 
the accuracy and the quantities $k_1h$ and $k_2h$ 
control the approximation interval. The functions 
(\ref{eq6.2}) possess the representation
\begin{equation}    \label{eq6.3}
v(r)=\frac{\phi(\ln r)}{r}, \quad
\phi(s)=\,h\sum_{k=k_1}^{k_2}\varphi(s-kh),
\end{equation}
in terms of the for $s$ going to infinity 
rapidly decaying window function
\begin{equation}    \label{eq6.4}
\varphi(s)=\exp(-\mathrm{e}^s+s).
\end{equation}
To check with which relative error the function
(\ref{eq6.2}) approximates $1/r$ on a given 
interval $1\leq r\leq R$, thus one has to examine 
how well the function $\phi$ approximates the 
constant $1$ on the interval $0\leq s\leq\ln R$. 
The functions (\ref{eq6.2}) decay exponentially 
as $r$ goes to infinity. This means in the given 
context that high frequencies in the Fourier 
representation of the functions under consideration 
are already inherently more or less strongly 
damped, without the need for additional measures. 

For $h=1$ and summation indices ranging from 
$k_1=-2$ to $k_2=50$, the relative error is, 
for instance, less than $0.0007$ on almost 
the whole interval $\mu\leq r\leq 10^{18}\mu$, 
that is, in the per mill range on an interval 
that spans eighteen orders of magnitude. 
Figure~3 shows the corresponding function 
$\phi$. The good approximation properties of 
the functions (\ref{eq6.2}) are underpinned 
by the analysis in 
\cite[Sect. 5]{Scholz-Yserentant} of the 
approximation properties of the corresponding 
infinite series. It has been shown there that 
these series approximate $1/r$ with a 
relative error
\begin{equation}    \label{eq6.5}
\sim 4\pi h^{-1/2}\mathrm{e}^{-\pi^2/h}
\end{equation}
as $h$ goes to zero, that is, already for 
surprisingly large $h$  with very high 
accuracy. Approximations of similar kind, 
which try to minimize the absolute instead 
of the relative error, have been studied by 
Braess and Hackbusch \cite{Braess-Hackbusch},
\cite{Braess-Hackbusch_2}.

\begin{figure}[t]   \label{fig4}
\includegraphics[width=0.93\textwidth]{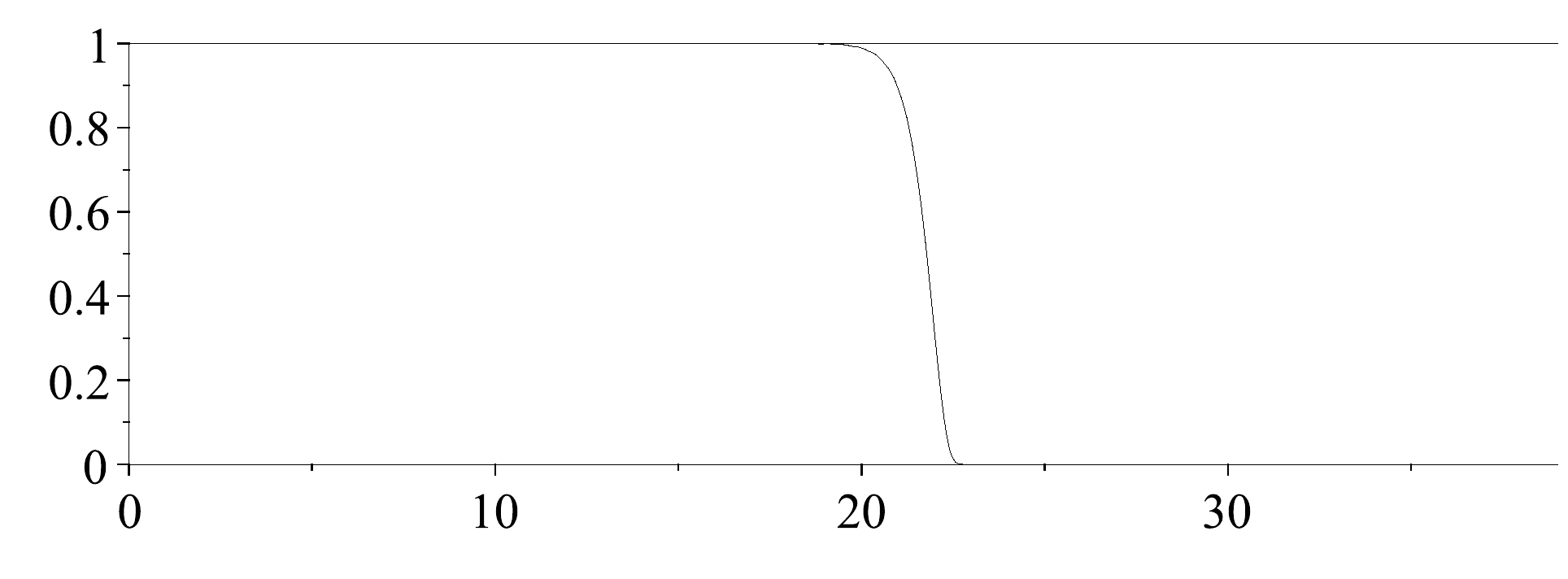}
\caption{The rescaled function $s\to\phi(s\ln 10)$ 
 approximating $1$ for $h=1$, $k_1=-2$, 
 and $k_2=50$.}
\end{figure}


\section{Symmetry properties of the solution 
 and of its approximations}
\label{sec7}

\setcounter{equation}{0}
\setcounter{theorem}{0}

Electronic wave functions are subject to the 
Pauli principle. Taking spin into account,
they are antisymmetric with respect to 
the exchange of the electrons, and if one 
considers the different spin components
of the wave functions separately, 
antisymmetric with respect to the exchange
of the positions of the electrons with the
same spin. The aim of this section is
to show that the given approximations
do not lead out of classes of functions 
with corresponding symmetry properties.

To every permutation $\pi$ of the electron 
indices $1,\ldots,N$, we assign two orthogonal 
matrices, first the $(m\times m)$-permutation 
matrix $P$ given by
\begin{equation}    \label{eq7.1}
Px|_i=x_{\pi(i)}, \quad i=1,\ldots,N,
\end{equation}
and then the much larger orthogonal 
$(n\times n)$-matrix $Q$ given by
\begin{equation}    \label{eq7.2}
Qy|_i=y_{\pi(i)}
\end{equation}
for the first $N$ subvectors 
$Qy|_i\in\mathbb{R}^3$ of $Qy$ and by
\begin{equation}    \label{eq7.3}
Qy|_{ij}=
\begin{cases}
\; y_{\pi(i),\pi(j)}, &\text{if $\pi(i)<\pi(j)$} \\
-y_{\pi(j),\pi(i)}, &\text{otherwise}
\end{cases} 
\end{equation}
for the remaining subvectors in $\mathbb{R}^3$ 
that are, as described in the introduction, 
labeled by the index pairs $(i,j)$, 
$i,j=1,\ldots,N$, $i<j$. Moreover, we assign 
to $Q$ the value
\begin{equation}    \label{eq7.4}
\epsilon(Q)=\mathrm{sign}(\pi),
\end{equation}
that is, the number $\epsilon(Q)=+1$, if $\pi$ 
is composed of an even number of transpositions, 
and the number $\epsilon(Q)=-1$ otherwise. 
By the definition (\ref{eq1.8}) of the 
matrix $T$, then 
\begin{equation}    \label{eq7.5}
 TP=QT
\end{equation}
holds. Let $G$ be a subgroup of the symmetric
group $S_N$, the group of the permutations of 
the indices $1,\ldots,N$. The matrices assigned 
to the elements of $G$ form then, with the 
matrix multiplication as composition, groups 
that are isomorphic to $G$.
We say that a function 
$u:\mathbb{R}^m\to\mathbb{R}$ is antisymmetric 
under the permutations in $G$, or shortly 
antisymmetric under $G$, if for all matrices 
$P=P(\pi)$ assigned to the $\pi\in G$
\begin{equation}    \label{eq7.6}
u(Px)=\mathrm{sign}(\pi)u(x)
\end{equation}
holds. We say that a function
$U:\mathbb{R}^n\to\mathbb{R}$ is antisymmetric 
under $G$ if for all these permutations $\pi$ 
and the assigned matrices $Q=Q(\pi)$ 
\begin{equation}    \label{eq7.7}
U(Qy)=\epsilon(Q)U(y)
\end{equation}
holds. Both properties correspond to
each other.

\begin{lemma}       \label{lm7.1}
If the function $U:\mathbb{R}^n\to\mathbb{R}$
is antisymmetric under $G$, so are its traces.
\end{lemma}

\begin{proof}
This follows by construction.
By (\ref{eq7.5}) we have
\begin{displaymath}
U(TPx)=U(QTx)=\epsilon(Q)U(Tx).
\end{displaymath}
Because of $\epsilon(Q)=\mathrm{sign}(\pi)$,
the proposition follows.
\qed
\end{proof}
We say that a function $U:\mathbb{R}^n\to\mathbb{R}$
is symmetric under the permutations in $G$, or
shortly symmetric under $G$, if for all matrices 
$Q=Q(\pi)$, $\pi\in G$,
\begin{equation}    \label{eq7.8}
U(Qy)=U(y)
\end{equation}
holds. Since the matrices $Q(\pi)$ assigned to the 
permutations in $\pi\in G$ are orthogonal,
rotationally symmetric functions have this property.

We show in the following that, for any
right-hand side $F$ of the equation (\ref{eq2.9}) 
that is antisymmetric under the permutations 
in $G$, also its solution (\ref{eq2.10}) is
antisymmetric under $G$ and that the same 
holds for the described approximations of this 
solution. 

\begin{lemma}       \label{lm7.2}
If the function $F\in W_0(\mathbb{R}^n)$ 
is antisymmetric under $G$, so are, for 
any measurable and bounded kernel $K$
that is symmetric under $G$, also the 
functions
\begin{equation}    \label{eq7.9} 
U(y)=\bigg(\frac{1}{\sqrt{2\pi}}\bigg)^n\!\int
K(\omega)\,\fourier{F}(\omega)\,
\mathrm{e}^{\,\mathrm{i}\,\omega\cdot y}\domega
\end{equation}
antisymmetric under the permutations 
in $G$.
\end{lemma}

\begin{proof}
From the orthogonality of $Q$, we obtain 
\begin{displaymath}    
U(Qy)=\bigg(\frac{1}{\sqrt{2\pi}}\bigg)^n\!\int
K(Q\omega)\,\fourier{F}(Q\omega)\,
\mathrm{e}^{\,\mathrm{i}\,\omega\cdot y}\domega
\end{displaymath}
and in the same way the representation
\begin{displaymath}    
\fourier{F}(Q\omega)=  
\bigg(\frac{1}{\sqrt{2\pi}}\bigg)^n\!\int
F(Qy)\,\mathrm{e}^{\,-\mathrm{i}\,\omega\cdot y}\dy.
\end{displaymath}
Because of $F(Qy)=\epsilon(Q)F(y)$, the Fourier 
transform of $F$ thus transforms like
\begin{displaymath}
\fourier{F}(Q\omega)=\epsilon(Q)\fourier{F}(\omega).
\end{displaymath}
As by assumption $K(Q\omega)=K(\omega)$, 
the proposition follows.
\qed
\end{proof}

The norm $\|\omega\|$ of the vectors $\omega$ in 
$\mathbb{R}^n$ is symmetric under the permutations 
in $G$, but also the norm of the vectors 
$T^t\omega$ in $\mathbb{R}^m$. This follows from 
the interplay of the matrices $P$, $Q$, and $T$, 
which finds expression in the relation 
(\ref{eq7.5}) and leads to
\begin{equation}    \label{eq7.10}
\|T^tQ\omega\|=\|P^tT^tQ\omega\|=
\|T^tQ^tQ\omega\|=\|T^t\omega\|.
\end{equation}
That is, every kernel $K(\omega)$ that depends 
only on the norms of $\omega$ and $T^t\omega$ 
is symmetric under $G$. Provided the right-hand 
side $F$ is antisymmetric under $G$, the same 
holds therefore for the solution (\ref{eq2.10})
of the equation (\ref{eq2.9}) and for all its
approximations calculated as described in the
previous sections. Additional smoothing of the 
right-hand side by means of kernels like those 
in (\ref{eq5.1}) does not change this picture.

To summarize, the solution of the equation 
(\ref{eq2.9}) and all its approximations 
and their traces completely inherit the 
symmetry properties of the right-hand 
side. We conclude that our theory is fully 
compatible with the Pauli principle. 
What is still missed, is a procedure for 
the recompression of the data between 
the single iteration steps, similar to 
that in \cite{Bachmayr-Dahmen} or
\cite{Dahmen-DeVore-Grasedyck-Sueli},
preserving the symmetry properties.


\bibliographystyle{spmpsci}
\bibliography{references}


\end{document}